# Movie Recommender System using critic consensus

A Nayan Varma
Department of CSE,
PES University,
Bangalore, India
nayanvarma3@gmail.com

Kedareshwara Petluri
Department of CSE,
PES University,
Bangalore, India
kevpetluri@gmail.com

*Abstract*—**Recommendation systems are perhaps one of the most important agents for industry growth through the modern Internet world. Previous approaches on recommendation systems include collaborative filtering and content based filtering recommendation systems. These 2 methods are disjointed in nature and require the continuous storage of user preferences for a better recommendation. To provide better integration of the two processes, we propose a hybrid recommendation system based on the integration of collaborative and content-based content, taking into account the top critic consensus and movie rating score. We would like to present a novel model that recommends movies based on the combination of user preferences and critical consensus scores.**

## I. INTRODUCTION (*HEADING 1*)

In the last 20 years, the Internet has transformed the way we live in all the aspects we can think about. From purchases, online streaming to news recommendations and emails, recommendation systems have become an indispensable part of our life that helps us to achieve the ultimate goal of making our decisions easier. Recommendation systems are systems that help us make our experience with an easy and attractive Internet. Since it helps us choose our shopping cart to decide which is the best news based on our preferences, plays an integral role in the life of our smartphone generation.

Movie recommendation systems help's us choose the best films we can like based on the user's previous interests and type of film. The goal of the recommendation system is delivering the movies that the user likes in the lowest possible number of suggestions, which helps the user to reduce time in navigation for the film. The main requirement for an effective recommendation system is that of the data. There is a data requirement and user preferences so that they can suggest the most precise recommendations. In general terms, recommendation systems can be classified into 2 types of recommendations based on content and collaborative recommendation systems.

During the design of the recommendation system, we have taken into account the need to integrate user preferences and critical consensus in the scoring mechanism. Our recommendations can provide better quality recommendations, since users preferences and critic scores take into account

## II. RELATED WORK

In the last 20 years, the Internet has transformed the way we live in all the aspects we can think about. From purchases, online streaming to news recommendations and emails, recommendation systems have become an indispensable part of our life that helps us to achieve the ultimate goal of making our decisions easier. Recommendation systems are systems that help us make our experience with an easy and attractive Internet. Since it helps us choose our shopping cart to decide which is the best news based on our preferences, plays an integral role in the life of our smartphone generation. We can broadly classify the work done into 2 categories - content Based Filtering and Collaborative Filtering. The details and trends for each of these methods is explained in this section.

### A. Content Based Filtering

Content based filtering can be defined as the recommendation based on the item types and the user profile description. Here we compute the similarity of different items and the items that the user wants based on the similarities between them. In the context of movies we take into instance the summary of the movies.

FLEX: A Content Based Movie Recommender [1] : In this paper the author uses a mixture of tf-idf method and doc2vec to create a hybrid model. Plots, regions , genre , year of release are used to help create a better and robust recommendation system.

Movie Recommender System Based on Percentage of View [2] : Generally we don't get an explicit feedback on most of the movies. Hence, the authors of these papers have proposed a unique method of rating and score calculation based on the number of views. The method is effective and has delivered better recommendations.

A content-based movie recommender system based on temporal user preferences [3] : In this paper the authors explore the recommendations based on the temporal user preferences. The model converts content attributes into Dirichlet Process Mixture Model to infer user preferences and provide a proper recommendation list.

Movie Recommender system using Sentiment Analysis [4] : Here the authors combine KNN methods with content based similarity using a weighted formula to compute a score and rate the movie accordingly.

### B. Collaborative Filtering

Collaborative filtering takes into account similarities between the users and the items while making recommendations.

There are mainly two algorithms constituting the memory based collaborative filtering which are the user-user collaborative filtering and item-item collaborative filtering. For user- user collaborative filtering, we find similar users and make recommendations to one user on the basis of what similar users liked whereas for item-item based collaborative filtering we recommend items similar to those liked by the user.

Optimizing Collaborative Filtering Recommender Systems [9] : This paper recognises the issues of assigning

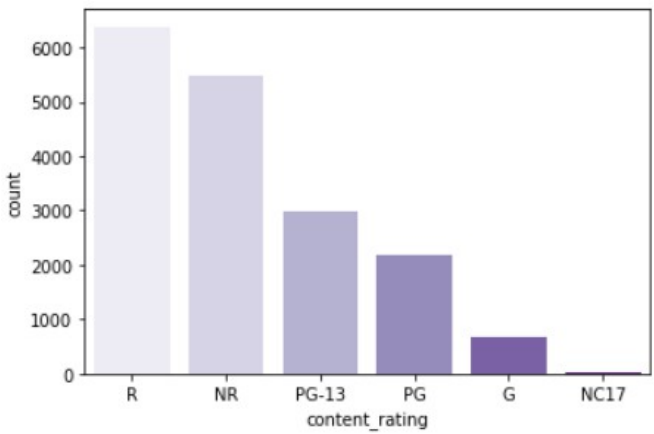
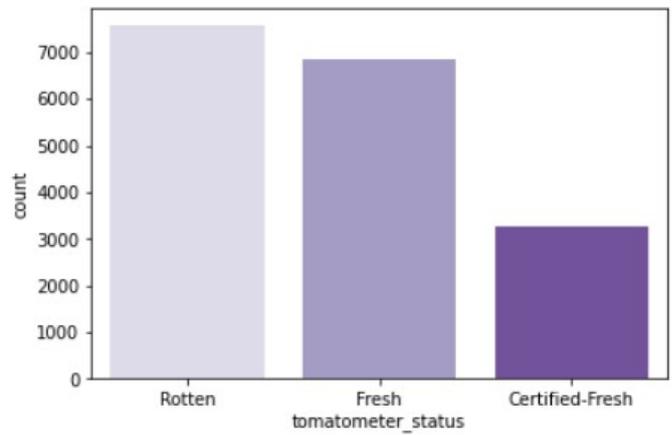

all items an equal weight and makes use of a GA based weighting technique in order to focus on the more useful ones and not give too much importance to the other. It has produced considerably better results.

An Optimization Method for Recommendation System Based on User Implicit Behavior [10] : In this paper, the issue of sparse matrix is dealt with. It is observed the that the matrix being sparse is due to the explicit behavior of the user is low. Hence, it also starts relying on the user's implicit behavior to remedy this issue. This include's whether the user chooses to watch a recommended video, time of watching a video and how frequently the user watches the video.

A Particle Swarm Approach to Collaborative Filtering based Recommender Systems through Fuzzy Features [8] : Here, the paper demonstrates the use of particle swarm optimization and the use of fuzzy sets in order to present the user information more efficiently.

measurement and others are deliberate, using entire proceedings, and not as an independent document. Please do not revise any of the current designations.

III.        DATA AND EDA

The datasets of Movielens and Rotten tomatoes have been merged with movie name and ID as the primary key. The RT dataset has a total of 7552 reviews by expert and top critics.The reviews have been given from various sources like the NYT, Hollywood reporter. The EDA Performed on the Rotten Tomato's dataset has been give below.

IV.        RESEARCH METHODOLOGY

A. Collaborative Filtering:

We use collaborative filtering in two different ways. In the first way we recommend movies to the user by finding users similar to the chosen user and recommending movies specifications that anticipate your paper as one part of the liked the similar users to the chosen user. For the second method, we find movies similar to those liked by the user and recommending these two users. The above mentioned methods are conjoined to build an effective movie recommender model. We find ways to improve this model performance and scalability wise in the methods discussed below.

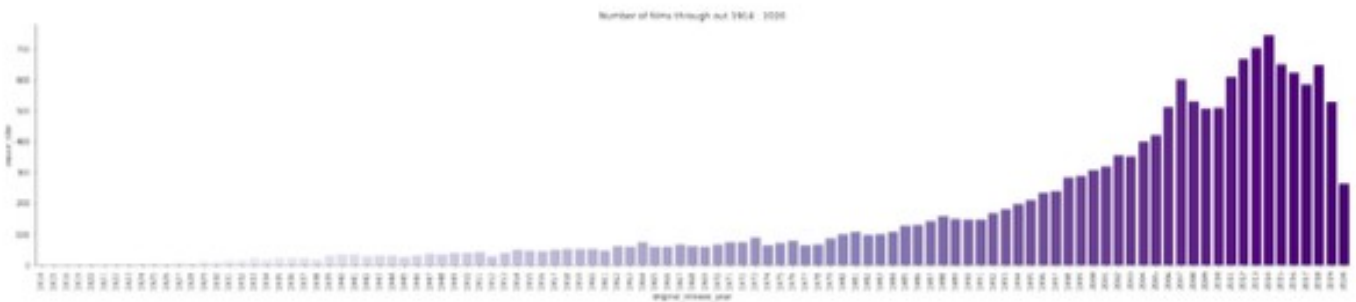

Fig. 3.  Number of films every year

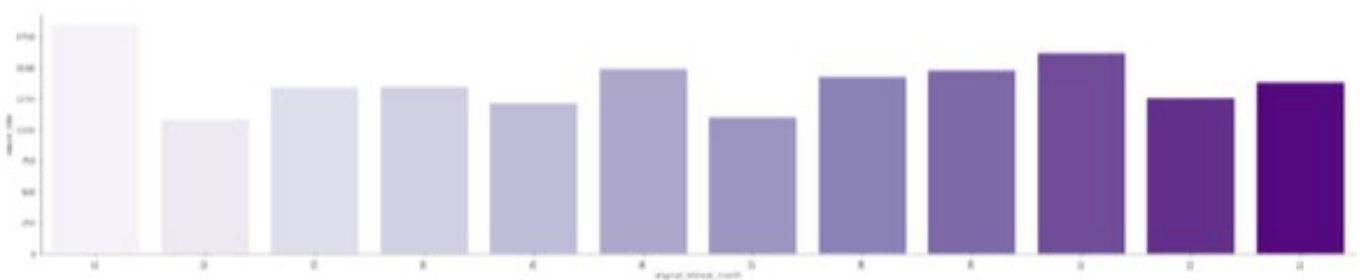

Fig. 4.  Movies released for each month

A Particle Swarm Approach to Collaborative Filtering based Recommender Systems through Fuzzy Features [8] : Here we use maintain user profiles in order to extract information from the implicit behaviour of the user. For instance the user rating a movie represents the how much he has enjoyed the movie. Further we take all this information and determine the similarity between the users and items with the help of fuzzy set using it later to rank our output.

Optimizing Collaborative filtering based recommendation systems [9] – Here we add in the feature of Genetic Algorithm based weight which helps filter out the choices which are less important and lay emphasis on the more important ones. This is an improvement over the original collaborative filtering as it reduce the amount of data to process in further computations.

An Optimization Method for Recommendation System Based on User Implicit Behaviour [10] : Here we take implicit information of the user such as whether he has watched the video, the amount of time he has watched the video in order to compensate for the sparsity of explicit information, and we make prediction which is later analysed using MAE.

### B. Content Based Filtering

The content based recommendation system is based on similarity between of a movie with another based on their summaries. There are many methods where the o effectively capture the the meaning of different sentences. The methods used here are SBERT and Universal sentence encoder.

*Sentence Bert:* Sentence-BERT [5] employments a Siamese arrange like engineering to supply 2 sentences as an input. These 2 sentences are at that point passed to BERT models and a pooling layer to create their embeddings. At that point utilise the embeddings for the combine of sentences as inputs to calculate the cosine similarity.

*Universal sentence encoder:* Universal sentence encoder [7] consists of transformer and deep averaging network. Sentences are periodically combined and best feature are extracted through softmax and in this way we get the best holistic representation of the sentences.

*Rating System:* A critic review to score model has been developed based on amazon movie reviews dataset. We have a top 1 accuracy of 83% and is effective for calculating the score. We have used RoBERTa based S-BERT [5] [6] model. This creates an effective model. The outputs produces were in range 0–5, and they are normalized in range of 0-0.2.

## V. Experiment

The first task was to calculate similar movie recommendations using collaborative filtering. We have used the methods of Swarm optimization and KNN to provide state-of-the-art results

Swarm optimization has been used to extract information about the user and fuzzy sets to calculate the similarity between users. We have taken into account user implicit behaviour to make up for the sparsity of user explicit behaviour such as whether user has chosen a video, the amount of time user has spent watching the video, how frequently the user watches the video.

KNN was used to extract similar users. Users based on similarity in movie preferences are segregated based on the principle of nearest neighbours. The intersection movies watched by similar users are recommended.

There are 3 similarity metrics available – jaccard, pearson and cosine - of which we will be using the pearson. We then make use of pairwise distances to calculate the pearson correlation coefficient with help of which a similarity matrix is obtained. With the help of this we can now assign scores to movies not assigned a score

The output of the collaborative filtering are the movie titles. We have used the following methods for preprocessing of the text

- Converted all words into lower case
- Used Demoji API to convert all emoji to words for effective representation
- Stop word removal
- Stemming
- Lemmatisation

We then pass these preprocessed movies title descriptions to SBERT or Universal sentence encoding. The encoding used are pre-trained state-of-the-art embeddings based on transfer learning.

We have used cosine similarity as a metric for calculating similarity. The outputs for each movie is cosine with description of movie's from the collaborative filtering. The critic score for each movie is added to its respective score.

Then based on the descending order of the calculated scores we compute the order of the movies. The computed movies are suggested based on their calculated weighted score. More the score better the recommendation.

In this way we are able to penalise the recommendation that is having low critic score yet keeping the essential issues of user preferences and context in light.

[('Grown Ups 2', 1.019047619047619),
('Psycho', 0.9114868760108947),
('Rebel Without a Cause', 0.9051149010658264),
('City Slickers', 0.9045331597328186),
('Lassie', 0.8879604458808898),
('Back to the Future', 0.8835068464279174),
('Grumpy Old Men', 0.8826252460479737),
('Thirteen', 0.8804184706122786),
('Garden State', 0.8706404907950039),
('A Nightmare on Elm Street 2: Freddy's Revenge', 0.8662821173667907),
('River's Edge', 0.8633230447769165),
('Crystal', 0.8626179814338684),
('Jason's Lyric', 0.8618192791938781),
('Bringing Out the Dead', 0.8575599850825937),
('Tom Jones', 0.8554827094078064)]

Fig. 5. Top 15 movies before critic score

[('Grown Ups 2', 1.0),
('Grumpy Old Men', 0.73262525),
('Fascination', 0.7250398),
('Rumble Fish', 0.7193358),
('The Family Man', 0.71772015),
('The Strangers', 0.7163406),
('Jumper', 0.71344244),
('Garden State', 0.7120198),
('Psycho', 0.7114869),
('Rebel Without a Cause', 0.7051149),
('A Guy Thing', 0.70479393),
('City Slickers', 0.70453316),
('Joe Dirt', 0.7035011),
('Thirteen', 0.6952333),
('The Majestic', 0.6893162)]

Fig. 6. Top 15 movie after critic score

## VI. Results and Discussion

### A. Qualitative Evaluation

The model has effectively been able to predict the various movies that the user likes and is able to penalize the movies with bad critic rating with less score. In this way we are able to holistically represent the movies that the user may watch only suggesting the best movies. The results have been displayed below. The base movie is chosen as " Grown Ups 2 ".

### B. Common Issues Evaluation

The following difficulties are always raised with recommender systems. We assess our system in light of these concerns and offer an implementation strategy to address them.

The New User Problem, for starters, is concerned with the situation in which a new user is added to the recommender system. He has yet to submit any ratings for any of the movies in the system. This is referred to as a User Cold Start. One simple option is to recommend top-rated movies or movies that have recently been introduced to this new user. Second, no new item is recommended, which is a source of concern. It's what's known as an Item Cold Start issue. When a new film is added to the system, it does not yet have any ratings attached to it. What is the best way to find it and recommend it? One resolution could be to recommend films in the same genre as the top-rated films. If the new film belongs to that genre, it will be noticed.However, in order to achieve this goal, we will need to create a system based on the film's genre.

## VII. Conclusion

Movie recommendation systems are very useful in our lives to help reduce time and efforts adopted to decide the value of the film. We have used state-of-the art methods like swarm based collaborative filtering, KNN with S-BERT and universal sentence encoder.This paper also includes how you can deal with the challenges to systems. Based on the results of the experiment, we can conclude that the system is effective for predicting high quality films.

### References

The template will number citations consecutively within brackets [1]. The sentence punctuation follows the bracket [2]. Refer simply to the reference number, as in [3]—do not use "Ref. [3]" or "reference [3]" except at the beginning of a sentence: "Reference [3] was the first ..."

Number footnotes separately in superscripts. Place the actual footnote at the bottom of the column in which it was cited. Do not put footnotes in the abstract or reference list. Use letters for table footnotes.

Unless there are six authors or more give all authors' names; do not use "et al.". Papers that have not been published, even if they have been submitted for publication, should be cited as "unpublished" [4]. Papers that have been accepted for publication should be cited as "in press" [5]. Capitalize only the first word in a paper title, except for proper nouns and element symbols.

For papers published in translation journals, please give the English citation first, followed by the original foreign-language citation [6].

1.  R. Singla, S. Gupta, A. Gupta, and D. K. Vishwakarma, "FLEX: A Content Based Movie Recommender," in *2020 International Conference for Emerging Technology (INCET)*, 2020, pp. 1–4. [Online]. Available: 10.1109/INCET49848.2020.9154163

2.  R. E. Nakhli, H. Moradi, and M. A. Sadeghi, "Movie Recommender System Based on Percentage of View," in *2019 5th Conference on Knowledge Based Engineering and Innovation (KBEI)*, 2019, pp. 656–660. [Online]. Available: 10.1109/KBEI.2019.8734976

3.  B. R. Cami, H. Hassanpour, and H. Mashayekhi, "A content-based movie recommender system based on temporal user preferences," in *2017 3rd Iranian Conference on Intelligent Systems and Signal Processing (ICSPIS)*, 2017, pp. 121–125. [Online]. Available: 10.1109/ ICSPIS.2017.8311601

4.  A. Chauhan, D. Nagar, and P. Chaudhary, "Movie Recommender system using Sentiment Analysis," in *2021 International Conference on Innovative Practices in Technology and Management (ICIPTM)*, 2021, pp. 190–193. [Online]. Available: 10. 1109/ ICIPTM52218.2021.9388340

5.  N. Reimers and I. Gurevych, "Sentence-BERT: Sentence Embeddings using Siamese BERT-Networks," 2019.

6.  Y. Liu, M. Ott, N. Goyal, J. Du, M. Joshi, D. Chen, O. Levy, M. Lewis, L. Zettlemoyer, and V. Stoyanov, "RoBERTa: A Robustly Optimized BERT Pretraining Approach," 2019.

7.  D. Cer, Y. Yang, S. yi Kong, N. Hua, N. Limtiaco, R. S. John, N. Constant, M. Guajardo-Cespedes, S. Yuan, C. Tar, Y.-H. Sung, B. Strope, and R. Kurzweil, "Universal Sentence Encoder," 2018.

8.  M. Wasid and V. Kant, "A Particle Swarm Approach to Collaborative Filtering based Recommender Systems through Fuzzy Features," *Procedia Computer Science*, vol. 54, pp. 440–448, 2015. [Online]. Available: https://doi.org/10.1016/j.procs.2015.06.051

9.  S.-H. Min and I. Han, "Optimizing Collaborative Filtering Recommender Systems," in *Advances in Web Intelligence*, P. S. Szczepaniak, J. Kacprzyk, and A. Niewiadomski, Eds. Berlin, Heidelberg: Springer Berlin Heidelberg, 2005, pp. 313–319.

10. P. Yi, C. Li, C. Yang, and M. Chen, "An Optimization Method for Recommendation System Based on User Implicit Behavior," in 2015 Fifth International Conference on Instrumentation and Measurement, Computer, Communication and Control (IMCCC), 2015, pp. 1537–1540. [Online]. Available: 10.1109/IMCCC.2015.326